%% file: main.tex
\documentclass[usenatbib]{mnras}

\usepackage{graphicx}
\usepackage{amsmath}
\usepackage{amssymb}
\usepackage{url}
\usepackage{CJKutf8}
\usepackage{array}
\usepackage{threeparttable}
\usepackage[T1]{fontenc}
\usepackage{orcidlink}

\input{new_commands}
\makeatletter
\newcommand*{\@rowstyle}{}
\newcommand*{\rowstyle}[1]{
\gdef\@rowstyle{#1}%
\@rowstyle\ignorespaces%
}
\newcolumntype{=}{
>{\gdef\@rowstyle{}}%
}
\newcolumntype{+}{
>{\@rowstyle}%
}
\newcolumntype{C}[1]{>{\centering\arraybackslash}p{#1}}
\makeatother

\pdfoutput=1

\title[Large-Volume EoR Simulations]{Thermal and Reionisation History within a Large-Volume Semi-Analytic Galaxy Formation Simulation} 

\author[Balu et al.]{Sreedhar Balu\orcidlink{0000-0002-5281-5151},$\!^{1, 2}$\thanks{E-mail:bsreedhar@student.unimelb.edu.au}
Bradley Greig\orcidlink{0000-0002-4085-2094},$\!^{1, 2}$
Yisheng Qiu\orcidlink{0000-0002-7716-1094},$\!^{3}$
Chris Power\orcidlink{0000-0002-4003-0904},$\!^{4,2}$
Yuxiang Qin\orcidlink{0000-0002-4314-1810},$\!^{1,2}$
\newauthor
Simon Mutch\orcidlink{0000-0002-3166-4614},$^{1, 2}$ and
J. Stuart B. Wyithe\orcidlink{0000-0001-7956-9758}$\!^{1, 2}$
\\
$^{1}$School of Physics, University of Melbourne, Parkville, VIC 3010, Australia\\
$^{2}$ARC Centre of Excellence for All Sky Astrophysics in 3 Dimensions (ASTRO 3D)\\
$^{3}$Institute for Astronomy, School of Physics, Zhejiang University, Hangzhou 310027, China\\
$^{4}$International Centre for Radio Astronomy Research, M468, University of Western Australia, 35 Stirling Hwy, Perth, WA 6009, Australia
}

\date{Accepted XXX. Received YYY; in original form ZZZ}

\pubyear{2022}

\begin{document}
\label{firstpage}
\pagerange{\pageref{firstpage}--\pageref{lastpage}}
\begin{CJK}{UTF8}{gkai} 
\maketitle
\end{CJK}

\begin{abstract}
\noindent
We predict the 21-cm global signal and power spectra during the Epoch of Reionisation using the \meraxes semi-analytic galaxy formation and reionisation model, updated to include X-ray heating and thermal evolution of the intergalactic medium. Studying the formation and evolution of galaxies together with the reionisation of cosmic hydrogen using semi-analytic models (such as $\meraxes$) requires N-body simulations within large volumes and high mass resolutions. For this, we use a simulation of side-length $210 \oneh$ Mpc with $4320^3$ particles resolving dark matter haloes to masses of $5\times10^8 \oneh \Msun$. To reach the mass resolution of atomically cooled galaxies, thought to be the dominant population contributing to reionisation, at $z=20$ of $\sim 2\times10^7 \oneh \Msun$, we augment this simulation using the \dforest Monte-Carlo merger tree algorithm (achieving an effective particle count of $\sim10^{12}$). Using this augmented simulation we explore the impact of mass resolution on the predicted reionisation history as well as the impact of X-ray heating on the 21-cm global signal and the 21-cm power spectra. We also explore the cosmic variance of 21-cm statistics within $70^{3}$ $h^{-3}$ Mpc$^3$ sub-volumes. We find that the midpoint of reionisation varies by $\Delta z\sim0.8$ and that the cosmic variance on the power spectrum is underestimated by a factor of $2-4$ at $k\sim 0.1-0.4$ Mpc$^{-1}$ due to the non-Gaussian nature of the 21-cm signal. To our knowledge, this work represents the first model of both reionisation and galaxy formation which resolves low-mass atomically cooled galaxies while simultaneously sampling sufficiently large scales necessary for exploring the effects of X-rays in the early Universe.
\end{abstract}


\section{Introduction}
The formation of the first luminous objects during the cosmic dawn resulted in the ionisation of the cosmic $\hone$ gas, rendering the intergalactic medium (IGM) transparent to UV photons. This period, termed the Epoch of Reionisation (EoR), constitutes the last major phase change of hydrogen in the Universe and had an impact on subsequent galaxy formation and evolution (\citealt{BarkanaReview}). A promising probe of this period is the 21-cm hyperfine spin-flip transition of $\hone$ which is sensitive to the evolution of the thermal and ionisation states of the IGM (\citealt{BibleReview}).

A number of low-frequency radio telescope arrays are in operation or are planned to detect this signal. Current instruments (MWA\footnote{\href{https://www.mwatelescope.org/}{www.mwatelescope.org}}, LOFAR\footnote{\href{https://www.astron.nl/telescopes/lofar}{www.astron.nl/telescopes/lofar}}, HERA\footnote{\href{http://reionization.org/}{reionization.org/}}) aim to detect the signal statistically via the 21-cm power spectrum (21-cm PS; \citealt{StuartReview}). While a detection has not yet been made, in recent years there has been significant progress in lowering the available upper limits (\citealt{LOFAR_limits}, \citealt{MWA_limits}, \citealt{HERA_limits}). In addition, the evolution of the all-sky averaged 21-cm global signal (21-cm GS) is being sought with experiments such as EDGES (\citealt{edges}) and SARAS (\citealt{SARAS3}). In the near future, the Square Kilometre Array (SKA; \citealt{SKAmain})\footnote{\href{https://www.skao.int/}{www.skao.int}} will provide an unprecedented ability to place observational constraints on the physics of this era by enabling the production of detailed 3-D 21-cm maps showing the distribution and evolution of the cosmic $\hone$.

For interpreting current and future observations it is important that realistic simulations of the early Universe are available and many authors have contributed to this effort (see \citealt{GnedinMadau} for a recent review). Simulations of the EoR are made challenging by the large range of scales involved. The main drivers controlling the ionisation and thermal states of the $\hone$ are respectively the intense UV and X-ray photons from star-forming galaxies (see \citealt{EoRbook} and references therein). X-ray photons have mean-free paths of the order of $10\rm s - 100\rm s$ of Mpc in the high-$z$ Universe, while the typical individual $\hii$ bubble sizes are $\sim 10 - 15$ Mpc (\citealt{Stu2004}, \citealt{Furlanetto2006}). It has also been shown that simulation volumes of sidelength $\gtrsim 100 \oneh$ Mpc are needed for convergent reionisation histories (\citealt{Iliev2014}) while $\gtrsim 200 \oneh$ Mpc are needed for convergent 21-cm PS (\citealt{Kim2016}, \citealt{Kaur2020}). These considerations necessitate simulations capable of resolving structures from a few Mpc in volumes of $\gtrsim$ 100s Mpc on a side.

At the same time, realistic EoR modelling requires the ability to resolve haloes down to at least the hydrogen cooling limit corresponding to a halo virial temperature of $T_\mathrm{vir}\sim10^4 \rm{K}$ and virial mass (\citealt{BarkanaReview})
\begin{equation}
    M_\mathrm{vir}(z)\sim4.4\times 10^{3}\bigg(\dfrac{T_{\rm vir}}{1+z}\bigg)^{3/2}h^{-1}\Msun
\end{equation}
These so-called atomically cooled haloes provide sites where gas  efficiently cools via atomic line transitions to form stars. Thus, to realistically simulate a representative volume of the early Universe, one requires large simulation volumes as well as sufficiently high mass resolutions.

Several techniques have been developed to simulate the EoR (\citealt{GnedinMadau}). Semi-numerical simulations (e.g. \citealt{Simfast21}, \citealt{21cmFAST}, \citealt{Maity2022}) typically associate ionising photon sources with the density peaks of evolved Gaussian random fields. As these models do not require running computationally expensive N-body simulations, they are able to achieve very large volumes (\citealt{Greig2022c}) as well as efficiently explore the available parameter space (e.g. \citealt{CMMC}). Their main drawback is the absence of detailed physics which self-consistently models a realistic galaxy population. On the other hand, achieving high resolution in large-volume hydrodynamical simulations is computationally expensive (see for example \citealt{CROC1}; \citealt{CoDaI}; \citealt{SPHINX}; \citealt{THESAN}). However, the computational overhead associated with hydrodynamical simulations precludes their use in parameter exploration. 

Semi-analytic models (SAMs; see \citealt{SomervilleReview} for a review) of galaxy formation ( e.g. \citealt{GalForm}, \citealt{Galacticus},  \citealt{SAGE}, \citealt{SAG}, \citealt{SHARK}) typically take merger trees from comparatively cheaper dark matter-only N-body simulations and evolve key baryonic components which describe the physical processes involved in galaxy formation, growth and evolution using simple but physically motivated prescriptions. Importantly, being based on N-body trees, the galaxies retain their association with the large-scale structure. These galaxy SAMs then provide a realistic galaxy population at a fraction of the cost of full hydrodynamical simulations. Coupling a galaxy SAM with a semi-numerical reionisation code can provide the best of both worlds: large-volume simulations of reionisation with a self-consistent, realistic population of galaxies. In this work, we use \meraxes (\citealt{Dragons3}), developed as part of the \textsc{DRAGONS} (Dark-ages Reionization And Galaxy formation Observables from Numerical Simulations) program, which couples a galaxy SAM model designed for galaxies in the high-$z$ Universe during the EoR with the semi-numerical code $\cmfast$ for simulating the reionisation process\footnote{A few other recent examples of SAMs incorporating reionisation calculations in the literature are \citealt{RSAGE}, \citealt{Visbal2020}, \citealt{Astraeus}.}. Additionally, for the first time, we implement the evolution of the neutral hydrogen gas spin temperature into $\meraxes$, taking into account heating by X-ray photons.

We run our updated \meraxes on a new dark matter-only N-body simulation which has a volume of $210^3$ $h^{-3}$ $\rm{Mpc}^3$ with $4320^3$ particles. This is the largest volume on which \meraxes has been deployed (previously $67.8^3$ $h^{-3}$ $\rm{Mpc}^3$; \citealt{Dragons19}). To achieve sufficient mass resolution (atomic cooling limit at $z=20$ of $\sim 2\times10^7 \oneh \Msun$) within our simulations we use \dforest -- a Monte Carlo algorithm-based code introduced in \cite{DARKFOREST}. This provides a unique dataset modelling both individual galaxy formation and evolution during reionisation in volumes large enough for exploring the effects of X-rays on the 21-cm signal from the cosmic dawn and the EoR. Importantly, this is the first time such a large volume coupled reionisation and galaxy SAM has been performed to study the 21-cm signal into the cosmic dawn. With our large volume, we are able to explore the impact of cosmic variance across the 21-cm statistics.

The paper is organised as follows: section \ref{sec:N-bodys} introduces the N-body simulations utilised in this work as well as its augmentation; section \ref{sec:meraxes} provides a brief summary of the \meraxes SAM and the calibration of its input model parameters. We analyse the resultant 21-cm signal from this model in section \ref{sec:sims_compare} and explore the cosmic variance across a broad range of statistics in section \ref{sec:cv}, and conclude in section \ref{sec:conclusion}. Our simulations use the best-fit parameters from the \cite{Planck2015}: $h=0.6751$, $\Omega_{\rm m}=0.3121$, $\Omega_{\rm b}=0.0490$, $\Omega_\Lambda=0.6879$, $\sigma_8=0.8150$, and $n_s=0.9653$. All quantities quoted are in comoving units unless otherwise stated.

\begin{figure}
		\includegraphics[width=\columnwidth ]{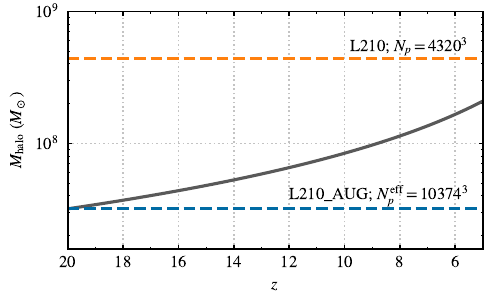}\vspace{-4mm}
		\caption{The mass of atomically cooled haloes (dark grey curve) as a function of redshift $z$ along with the representative halo mass resolution of the \gensim (orange dashed) and \genpsim (blue dashed) simulations.}
		\label{fig:mass_res}
\end{figure}

\section{N-body simulations and their augmentation}\label{sec:N-bodys}

In this section, we introduce the N-body simulation used in this work as well as an outline of the augmentation pipeline.

\begin{figure*}
		\includegraphics[width=\textwidth]{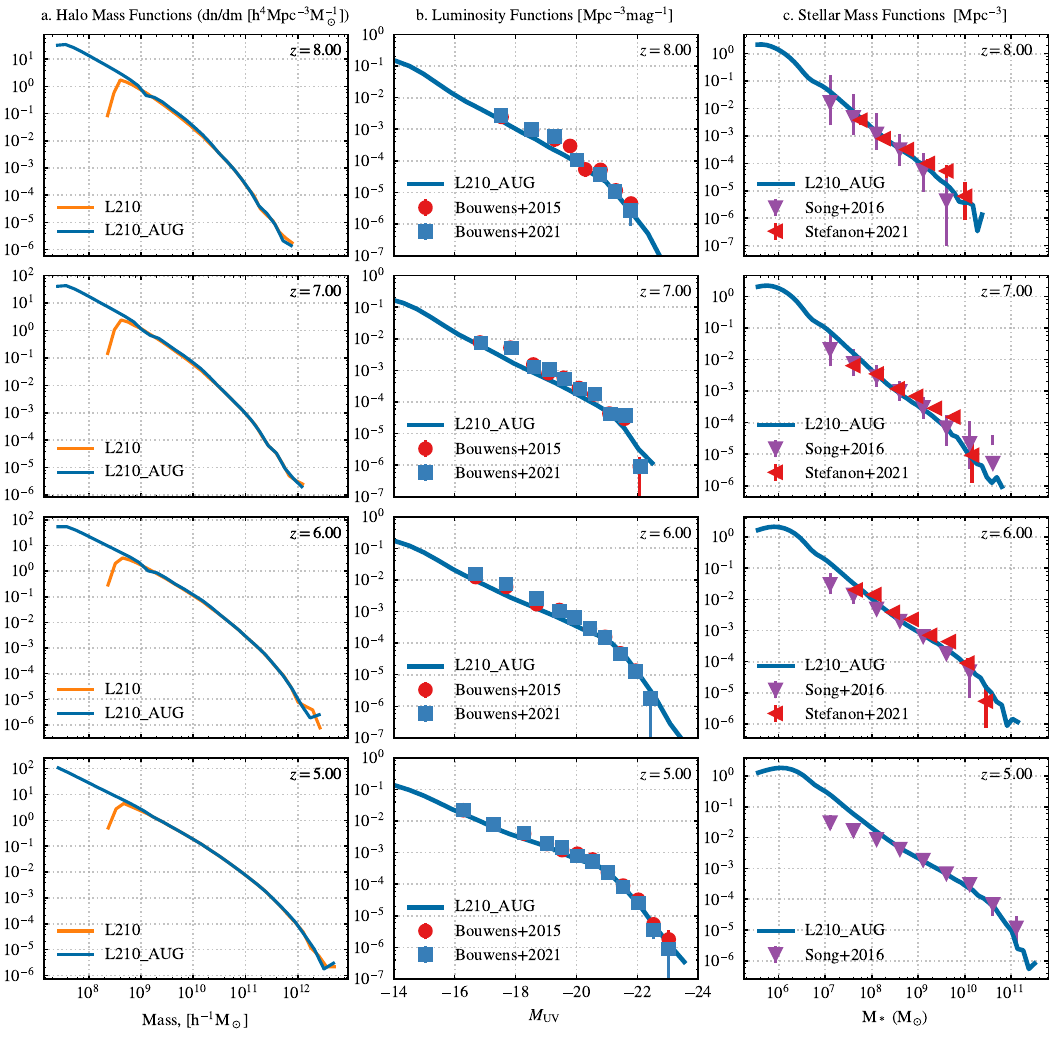}\vspace{-4mm}
		\caption{\label{fig:calibration}We show the HMF (first column), the LF (middle column), and SMF (right column) at $z\sim$ 8, 7, 6, \& 5 for both $\gensim$ (orange curve) and $\genpsim$ (blue curve). The first column shows the impact of the augmentation, highlighting the mass of the smallest haloes that are resolved. The final two columns demonstrate the calibration of the $\meraxes$ using existing data. For the LF we use \citet{Bouwens2015} \& \citet{Bouwens2021}, and for the SMF we use \citet{Song2016} \& \citet{Stefanon2021}.
}
\end{figure*}

\begin{table*}
	\centering
    \begin{tabular}{cccc}
    \hline 
    \vspace{-2mm}\\
    Name & Mass resolution & $L_{X<\text{2 keV}}/\mathrm{SFR}$ & Comments \\ 
     &  $[h^{-1} \Msun]$ &  [erg $\rm s^{-1} \Msun^{-1} \rm{yr}$] &  \\ \vspace{-2mm}\\
    \hline
    \vspace{-2mm}\\
     
    \gensim & $2.98 \times 10^8$ & $3.16 \times 10^{40}$ & Fiducial simulation \\    
    
    \genpsim & $2.12 \times 10^7 $ & $3.16 \times 10^{40}$ & Augmented fiducial simulation.\\
    
    \mgenpsim  & $2.12 \times 10^7 $ & $3.16 \times 10^{38}$ & Same as \genpsim but with $1/100$th of the galaxy X-ray luminosity\\
    
    \pgenpsim  & $2.12 \times 10^7 $ & $3.16 \times 10^{42}$ & Same as \genpsim but with $100 \times$ of the galaxy X-ray luminosity \\
    \vspace{-2mm}\\
    \hline
    \vspace{-2mm}\\
    \textsc{L210\_nr} & $2.98 \times 10^8$ & $3.16 \times 10^{40}$ & \gensim without recombinations \\ 
    
    \textsc{L210\_AUG\_nr} & $2.12 \times 10^7 $ & $3.16 \times 10^{40}$ & \genpsim without recombinations\\
    \vspace{-2mm}\\
    \hline 
    \end{tabular}
    \caption{The names, mass resolution, X-ray luminosity and a brief description of all the simulations used in this work.}
    \label{table:sims}
\end{table*}

\subsection{\gensim Simulation}

We use the L210\_N4320 (hereafter $\gensim$) box of the \textsc{Genesis} suite of N-body simulations (Power et al. in prep). This simulation is $210 \oneh$ Mpc on a side and  consists of $4320^3$ dark matter particles of mass $m_{\rm p} = 9.95 \times 10^{6} \oneh \Msun$. The halo mass resolution is $\sim 5 \times 10^8 \oneh \Msun$ based on a minimum of 50 particles. The simulation was evolved from $z=99$ down to $z=5$ using the \textsc{SWIFT} code (\citealt{SWIFT}) and the haloes were identified via friends-of-friends by the \textsc{VELOCIraptor} halo-finder (\citealt{VELOCIRAPTOR}). Halo catalogues are saved over 120 snapshots evenly distributed in dynamical time between redshifts 30 and 5. The merger trees were generated using \textsc{TREEFROG} (\citealt{TREEFROG}).

\subsection{\genpsim Simulation}
To increase the mass resolution of the \gensim simulation from $\sim 5 \times 10^8 \oneh \Msun$ to the atomic hydrogen cooling limit at $z=20$ ($\sim 2\times10^7 \oneh \Msun$) we augment it by extending the merger trees to lower mass haloes. This is achieved using $\dforest$ (\citealt{DARKFOREST}), a Monte-Carlo (MC) based algorithm which we summarise below. We call this new simulation $\genpsim$, which provides a unique dataset for exploring galaxy formation physics and its impact on the timing and morphology of the EoR. Fig. \ref{fig:mass_res} shows the mass resolution of both \genpsim (orange dashed) and \gensim (blue dashed) along with the atomic cooling limit (dark grey curve) for the relevant redshifts. We point out that the augmentation algorithm works backward in time (in our case from $z=5$).

\dforest uses an updated prescription of \cite{Benson2016} for augmenting merger trees and works on what are termed ``simple branches'' -- merger tree branches that are composed of a halo and all of its immediate progenitors. To add new haloes to the existing merger trees new simple branches are  generated using the algorithm outlined in \cite{Parkinson2008} which employs a conditional mass function, with extra parametrisation (to take care of the differences between the analytic halo mass functions (HMFs) and the ones from N-body simulations) derived from the Extended Press Schechter theory (\citealt{EPS3}, \citealt{EPS2}, \citealt{EPS1}). Each halo is split into two (binary splits) in small internal time steps: we choose these time steps, $dz_1$, such that $|dz_1| << z_1/z_2$ where $z_1$ is the redshift of the halo and $z_2$ is the redshift of its immediate progenitors. This construction is repeated until we have an MC merger history. These new MC branches, by construction, have a higher mass resolution than the N-body trees. Building on  the methods employed in \cite{Benson2016}, the new branches are used to augment the existing N-body merger tree.

For this we first define a mass threshold, $M_{\rm cut}$, which serves as a dynamic boundary between the N-body and MC halo populations in the final augmented merger tree thus helping us to ``average out'' the differences between these two populations. If all the newly added haloes in the generated MC simple branch are less massive than $M_{\rm cut}$ then those haloes are  attached to the original N-body simple branch. As a result, the augmented simple branch will have both the MC haloes for $M_{\rm halo}<M_{\rm cut}$ and the original N-body haloes for $M_{\rm halo}>M_{\rm cut}$. The final augmented merger tree with these MC branches grafted onto it will thus have both N-body as well as MC haloes with the $M_{\rm cut}$ serving as the barrier separating the N-body and MC haloes. The resultant ``hybrid'' merger tree will have the same mass resolution as the MC simple branches. \cite{Benson2016} used a constant value for $M_{\rm cut }$. We allow $M_{\rm cut }$ to take values $\in [M_{\rm cut}^{\rm min}, M_{\rm cut}^{\rm max}]$. For every simple branch, the augmentation starts with $M_{\rm cut}=M_{\rm cut}^{\rm min}$, but incrementally increases it if the MC simple branch is not deemed fit\footnote{This can happen for instance, if the number of MC haloes are less than that of the N-body haloes and/or if the difference in mass of the MC haloes of $M_{\rm halo}>M_{\rm cut}$ and the corresponding N-body haloes in the simple branch are larger than a precision parameter. Note that these MC haloes (with masses above $M_{\rm cut}$) do not end up in the final merger tree but are used solely as a check on the augmentation algorithm. See \cite{DARKFOREST} for further details.} to be attached to the N-body simple branch. \cite{DARKFOREST} used $[M_{\rm cut}^{\rm min}, M_{\rm cut}^{\rm max}] \equiv [100 m_{\rm p}, 2500 m_{\rm p}]$ where $m_{\rm p}$ is the particle mass of the unaugmented simulation. Instead, we use a $M_{\rm cut}^{\rm min} = 120 m_{\rm p}$ taking into account the higher resolution of the \gensim simulation while $M_{\rm cut}^{\rm max}$ remains unchanged. This choice was made so that we can retain a larger fraction of the N-body haloes in the \genpsim simulation. The mass resolution of the augmented simulation is a free parameter of $\dforest$ which we have chosen to be the atomic cooling limit at $z \sim 20$.

In the first column of Fig. \ref{fig:calibration} we show the HMFs of the augmented (\genpsim in {blue}) and unaugmented ($\gensim$ in orange) boxes at $z =$ 8, 7, 6, and 5. There is a turn-over of the HMF of \gensim prior to the resolution limit because of the incomplete identification of haloes by the halo-finder. This further motivates the need for augmentation to obtain all the haloes down to the desired mass limit. The HMFs of \genpsim extend out to the desired mass resolution with the smallest haloes resolved in the augmented tree having a mass of $\sim 2 \times 10^7$ $\oneh$ $\Msun$. 

Since reionisation morphology depends on the location of photon sources, it is important that the positions of the MC haloes are assigned appropriately.  $\dforest$ determines the positions and assigns velocities to the newly added MC haloes. We apply the non-linear halo bias prescription from \cite{Ahn2015} on the input dark matter density field from $\gensim$ to generate a halo density field. This is normalised and used as a one-dimensional probability distribution from which the MC haloes are assigned their positions by random sampling. The MC haloes are placed uniformly within a voxel in such a way that the number of haloes follows a Poisson distribution. The accuracy of this random sampling method (which we assert by comparing the 2-point correlation functions between the MC and N-body haloes in the same mass ranges) has been shown to depend on the grid size. We performed a convergence test to determine the resolution providing the best performance and use $512^3$ cells for our calculations. This is also partly motivated by the resulting grid size of $0.4~h^{-1}$ Mpc being smaller than the $\hii$ bubble sizes (\citealt{Furlanetto2006}). This method is used to assign positions to every new MC that is not a progenitor of another MC halo (i.e. it has just been resolved for the first time).

The evolution of the MC haloes' position with time is based on their peculiar velocity field, $\boldsymbol{v}(\boldsymbol{x},t)$, using the linear continuity equation as
\begin{equation}
    \nabla \boldsymbol{v}(\boldsymbol{x},t) = -\frac{1}{\Delta t}[\mathcal{D}(\boldsymbol{x}, t_1) - \mathcal{D}(\boldsymbol{x}, t_2)],
\end{equation} 
where $\mathcal{D}(\boldsymbol{x}, t) = b(M,t)\delta_{DM}(\boldsymbol{x},t)$ is the halo density field with $b(M,t)$ the linear halo bias (\citealt{Tinker2010}) and $\delta_{DM}(\boldsymbol{x}, t)$ is dark-matter overdensity field, and $\Delta t$ is the time-step between the snapshots. Once again, we find that the choice of grid sizes for determining the halo density fields affects the accuracy of the halo positions. \cite{DARKFOREST} used an evolving (with $z$) grid resolution. Based on their results, we use a $256^3$ grid resolution at $z = 5-6$,  $64^3$ at $z = 6-8$, and $32^3$ grids at $z>8$ after compensating for the differences in the simulation volumes.

As detailed in \cite{DARKFOREST} we run a number of tests to ensure that the MC haloes are introduced without compromising the accuracy of the underlying \gensim simulation. Specifically, while evolving the position of the MC haloes the 2-point correlation of the halo positions and the velocity distribution of the haloes are checked to ensure they are consistent. The interested reader is referred to \cite{DARKFOREST} for a detailed explanation of the augmentation algorithm.

\section{Modelling Galaxies and the Epoch of Reionisation}\label{sec:meraxes}

\subsection{Galaxy formation using \meraxes}

SAMs enable fast and efficient modelling of galaxies and their properties within cosmological volumes. In this work, we use the \meraxes (\citealt{Dragons3}) SAM which was specifically designed to study the interplay and feedback between galaxy formation and evolution, and reionisation. Since its introduction \meraxes has undergone several updates. These include AGN feedback (\citealt{Dragons10}) as well as updates to supernova feedback, recycling and chemical enrichment of the ISM, and reincorporation of the ejected gas (\citealt{Dragons19}). 

\meraxes includes detailed, physically motivated prescriptions for processes including baryonic infall into a dark matter halo, radiative cooling of this infalling gas, star formation, supernova feedback which can heat up the cold gas, mass recycling whereby the ejected material from a supernova can participate in star formation again, metal enrichment of the interstellar medium (ISM), and reincorporation of the gas that is ejected from the galaxy but still bound to the dark matter halo. The dynamical time of a typical galactic disc at high redshift is $\sim 10$s of Myr (which is similar to the lifetime of massive stars). Our N-body simulations have therefore been constructed with high cadence (with a mean value of 10 Myr between $z\sim 30-5$) and \meraxes also includes time-dependent feedback based on the star formation history.

At each snapshot, the baryonic content of a dark matter halo increases up to $f_b f_{\rm mod}M_{\rm vir}$ in the form of pristine primordial gas. Here, $f_b = \Omega_{\rm b}/\Omega_{\rm m}$ is the baryon fraction of the Universe and $f_{\rm mod}$ is the baryon fraction modifier which couples the feedback of reionisation to galaxy formation. This newly acquired baryonic gas is deposited into a shock-heated and quasi-static hot-gas reservoir of the galaxy. The fraction of this hot gas which has a cooling time less than the dynamical time of the halo cools radiatively to a much colder gas cloud. This cold gas then participates in star formation following the \cite{Kauffmann1996} model. The cold gas reservoir can also be depleted by the feedback from supernova and active galactic nuclei (\citealt{Dragons10}). 

\cite{Dragons19} introduced a dust model into \meraxes facilitating the computation of dust attenuated luminosity functions (LFs). The implementation is based on a dust attenuation model from \cite{Charlot2000}. Within a Bayesian framework, \cite{Dragons19} explored three parameterisations for dust in \meraxes linked to the SFR, dust-to-gas (DTG) ratio, and gas column density (GCD). In this work, we use the DTG model which depends on the cold gas' metallicity and mass.

To model reionisation and investigate the role of photoionisation feedback on the high-$z$ galaxies, $\meraxes$ includes a modified version of $\cmfast$ (described in the next section; \citealt{21cmFAST}, \citealt{21cmFASTv3}). At each snapshot, once the galaxies are identified and all the associated gas reservoirs are updated appropriately, \meraxes models their impact on the $\hone$ in the IGM.

\subsection{Reionisation in \meraxes}

Reionisation is incorporated self-consistently in \meraxes using the computationally efficient semi-numerical code $\cmfast$. Using perturbation theory, \cmfast generates evolved density and velocity fields which are then converted to stellar mass and star formation rate (SFR) grids using a simple parameterisation to describe the galaxies. In \meraxes the first two fields come directly from the N-body simulations thus retaining the non-linear effects of structure formation while the stellar mass grids are computed realistically by \meraxes making use of the full galaxy properties. In this work, we extend the reionisation calculations of \meraxes to additionally follow the evolution of the spin temperature, $T_{\rm S}$, of $\hone$ by incorporating the heating and ionisation of the IGM by X-rays following the same approach taken within $\cmfast$. For all of the reionisation calculations we use a grid resolution of $1024^3$ corresponding to a cell resolution of $\sim 0.2$ $\oneh$ Mpc, which is smaller than the typical size of $\hii$ regions during the EoR (\citealt{Stu2004}). In this section, we describe the implementation of reionisation and thermal evolution in $\meraxes$.

\subsubsection{$\hi$ reionisation}

The ionisation state of the IGM is determined directly from the stellar mass grids following the excursion-set formalism (\citealt{Furlanetto2004}). Here, the total integrated number of ionising photons is compared to the number of neutral atoms plus recombinations within spheres of radius $R$, centred at location $\boldsymbol{x}$ and redshift $z$. A simulation cell is flagged as ionised if
\begin{equation}\label{eq:excursion}
    N_{\rm b*} (\boldsymbol{x}, z | R) N_\gamma f_{\rm esc} \geq  N_{\rm atom} (\boldsymbol{x}, z | R) (1 + \bar{n}_{\rm rec}) (1 - \bar{x}_e),
\end{equation}
where $N_{\rm b*} (\boldsymbol{x}, z | R)$ is the number of stellar baryons in the sphere, $N_\gamma$ is the average number of ionising photons per stellar baryons, and $f_{\rm esc}$ is the escape fraction of the photons.  $N_{\rm atom} (\boldsymbol{x}, z | R)$ is the total number of baryons in the same volume, and $(1 - \bar{x}_e)$ accounts for secondary ionisations caused by the X-ray photons. \cite{Sobacchi2014} have shown that recombinations inside Lyman limit systems can significantly reduce the sizes of $\hii$ regions. Following their implementation in $\cmfast$, through a sub-grid prescription, we account for recombinations via the $\overline{n}_{\rm rec}$ term which is the mean number of recombinations.
We decrease $R$ from a maximum of 50 Mpc, which is the mean-free path in the IGM post-reionisation (\citealt{Songaila2010}, \citealt{Becker2021}), down to the size of a voxel, $R_{\rm cell}$. 

The local ionisation state of the IGM is used to evaluate the value of $f_{\rm mod}$ for all the galaxies in the volume. The amount of fresh gas accreted in the next snapshot by the host haloes of the galaxies is then suppressed by a factor of $f_{\rm mod}$ thus enabling \meraxes to couple galaxy evolution with reionisation. This gives us a reionisation scenario that is self-consistent and regulated by the UV background (UVB). An exploration of the interplay between the galaxies and reionisation and its impact on the 21-cm PS (though only in the post-heating regime) with \meraxes is given in \cite{Dragons5}.

\subsubsection{Spin temperature field}
    
The 21-cm signal depends upon the spin temperature $T_{\rm S}$ which quantifies the population ratio of the two $\hone$ hyperfine energy levels. $T_{\rm S}$ is sensitive to the thermal state of the IGM which is influenced by the X-ray photons and is given by
\begin{equation}
    T_{S}^{-1} = \dfrac{T_{\rm CMB}^{-1} + x_{\alpha} T_\alpha^{-1} + x_{c} T_{\rm{K}}^{-1}}{1 + x_{\alpha} + x_{\rm{c}}},
\end{equation}
where $T_{\rm CMB}$, $T_{\alpha}$, and $T_{\rm K}$ are the CMB, colour, and gas kinetic temperatures respectively, $x_\alpha$ is the Wouthuysen-Field (WF) coupling constant (\citealt{Wouthuysen1952} \& \citealt{Field1958}) and $x_{\rm c}$ is the collisional coupling coefficient. We take $T_\alpha = T_{\rm K}$, and $x_c$ is computed as 
\begin{equation}
    x_c = \dfrac{0.0628~\rm{K}}{A_{10}T_\gamma} \sum_i n_i\kappa^{iH}(T_{\rm K}),
\end{equation}
where $A_{10}$ is the Einstein spontaneous emission coefficient, and $i\in\{\hone, \rm{~free~electrons}~(e), \rm{~free~protons}~(p)\}$ and the $\kappa$'s refers to the corresponding collisional coefficients. We compute
$x_\alpha$ as
\begin{equation}
x_\alpha=1.7\times10^{11}(1+z)^{-1}S\alpha J_\alpha,
\end{equation}
where $S_\alpha$ is an order-of-unity correction factor involving atomic physics and $J_\alpha$ (pcm$^{-2}$ s$^{-1}$ Hz$^{-1}$ sr$^{-1}$ where ``p'' denotes proper units) is the $\lya$ background flux. We follow the method outlined in \cite{21cmFAST} for computing $S_\alpha$ and $J_\alpha$.

Following $\cmfast$, we compute the gas kinetic temperature $T_{\rm K}$ and the ionised fraction $x_{\rm e}$ at position $\boldsymbol{x}$ and redshift $z$ as:

\begin{equation}
\dfrac{\mathrm{d}x_{e}(\boldsymbol{x}, z)}{\mathrm{d}z} = \dfrac{\mathrm{d}t}{\mathrm{d}z}[\Gamma_{\mathrm ion} - \alpha_{\mathrm A}Cx_{e}^{2} n_{b}f_{\mathrm H}],
\end{equation}
\begin{equation}\label{eq:T_k}
\begin{split}
    \dfrac{\mathrm{d}T_{\mathrm K}(\boldsymbol{x},z)}{\mathrm{d}z} = \dfrac{2}{3k_{\rm B} (1+x_{e})}\dfrac{\mathrm{d}t}{\mathrm{d}z}\sum_p\epsilon_p & +  \dfrac{2T_K}{3n_b}\dfrac{\mathrm{d}n_b}{\mathrm{d}z}\\ & - \dfrac{T_K}{1+x_e}\dfrac{\mathrm{d}x_e}{\mathrm{d}z},
\end{split}
\end{equation}
where $n_b = \overline{n}_{b,0}(1+z)^3[1+\delta_{nl}(\boldsymbol{x},z)]$ is the total baryonic number density (H+He), $\epsilon_p(\boldsymbol{x},z)$ is the heating rate per baryon for process $p$ (in erg $\rm s^{-1}$), $\Gamma_{\rm ion}$ is the ionisation rate per baryon, $\alpha_A$ is the case-A recombination coefficient, $C$ is the clumping factor on the scale of the simulation cells ($C \equiv \langle n^2\rangle/ \langle n \rangle^2 = 2$; \citealt{Sobacchi2014}), $k_B$ is the Boltzmann constant, and $f_H$ is the hydrogen number fraction. Equation \ref{eq:T_k} governs the thermal evolution of the gas and incorporates contributions from Compton heating (first term), adiabatic cooling and heating due to Hubble expansion and structure formation respectively (second term), and the change in internal energy of the system due to the changing number of particles (third term).

Both $\Gamma_{\rm ion}$ and $\epsilon_{p}$ depend on the angle-averaged specific X-ray intensity $J(\boldsymbol{x}, E, z)$. For a voxel with location $\boldsymbol{x}$ at redshift $z$, the X-ray intensity at energy $E$,  $J(\boldsymbol{x}, E, z)$, is computed by integrating the comoving X-ray specific emissivity $\epsilon_X(\boldsymbol{x}, E_e, z^{'})$ back along the light-cone as

\begin{equation}
    J(\boldsymbol{x}, E, z) = \dfrac{(1+z)^3}{4\pi} \int_z^\infty dz' \frac{cdt}{dz^{'}} \epsilon_X e^{-\tau},
\end{equation}
where $e^{-\tau}$ accounts for the attenuation of the X-ray photons by the IGM i.e. the probability that an X-ray photon emitted at redshift $z^{'}$ survives till $z$.

We relate the comoving X-ray specific emissivity $\epsilon_X(\boldsymbol{x}, E_e, z^{'})$, evaluated in the emitted frame where $E_e = E (1 + z^{'}) / (1 + z)$, to the star formation rate density SFRD$(\boldsymbol{x}, E_e, z^{'})$ in the voxel 

\begin{equation}
\epsilon_X(\boldsymbol{x}, E_e, z') = \dfrac{L_X}{\mathrm{SFR}}\times \mathrm{SFRD}(\boldsymbol{x}, E_e,  z'),
\end{equation}
where $L_X/\mathrm{SFR}$ is the specific X-ray luminosity per unit star formation that is capable of escaping the galaxy in units of  [erg $\rm s^{-1} \Msun^{-1} \rm{yr}$]. 

Unlike $\cmfast$, where the SFRD is calculated from the density field and collapse fraction, we compute the SFRD directly from \meraxes making use of our galaxies' properties. Feedback from reionisation can thus alter the star formation rates of galaxies. $L_X/\mathrm{SFR}$ is assumed to follow a power-law of the form $L_X/\mathrm{SFR} \propto E^{-\alpha_X}$ where $E$ is the photon energy and is normalised with respect to the soft-band X-ray luminosity as 
\begin{equation}\label{eq:e_x}
L_{X<\text{2 keV}}/\mathrm{SFR} = \int_{E_0}^{\text{2 keV}} dE_e \text{ }L_X/\mathrm{SFR}.
\end{equation}
Here, $E_0$ is a threshold energy that fixes the lowest energy of X-ray photon capable of escaping the galaxy by producing a sharp cut-off in the X-ray spectrum that accounts for where the X-rays are absorbed by the high column density gas within the galaxy.

We thus have three free parameters characterising the X-ray properties of the galaxies: $L_{X<\text{2 keV}}/\mathrm{SFR}$, $E_0$, and $\alpha_X$. In this work we only vary $L_{X<\text{2 keV}}/\mathrm{SFR}$ keeping the other two fixed\footnote{See \cite{GreigCMMC2} for an exploration of $E_0$ and $\alpha_X$. The reader is also cautioned that a direct comparison with this work is not straightforward since they do not have a realistic galaxy population.}. We set $\alpha_X = 1$ consistent with the observations of high-mass X-ray binaries in the local Universe (\citealt{Mineo2012},  \citealt{Fragos2013}, \citealt{Pacucci2014}). Motivated by \cite{Das2017} we adopt a value of $E_0 = 0.5 \mathrm{\text{ keV}}$ throughout this work. The physical interpretation of the upper limit of $2 \mathrm{\text{ keV}}$ in the integral of equation (\ref{eq:e_x}) is that X-ray photons with higher energies have mean-free paths longer than the Hubble length and thus do not interact with the IGM.

\subsubsection{Brightness temperature field}
The 21-cm brightness temperature field is given by (\citealt{BibleReview}):
\begin{equation}
\label{eq:T_b}
\begin{split}
\delta T_{b} (\nu)  & = \dfrac{T_{S} - T_{\gamma}}{1 + z} (1 - e^{-\tau_{\nu_0}})\\
&\approx  27 x_{\textsc{\ion{H}{i}}}(1 + \delta_{nl} ) \left( \dfrac{H}{dv_{r}/dr + H} \right) \left( 1 - \frac{T_{\gamma}}{T_{S}} \right)\\
&\mathrm{\hspace{0.3cm}}\times \left( \frac{1+z}{10} \frac{0.15}{\Omega_{M} h^{2}}\right) \left(\dfrac{\Omega_{b} h^{2}}{0.023}\right)
\mathrm{ mK},
\end{split}
\end{equation}
where $T_\gamma$ is the background radiation (usually assumed to be the CMB) impinging upon the $\hone$ cloud, $\tau_{\nu_0}$ is the optical depth at the 21-cm transition frequency $\nu_0$, $1 + \delta_{nl}$ is the density contrast in the dark matter field ($\delta_{nl} = \rho/\bar \rho - 1$), $H(z)$ is the Hubble parameter at the redshift $z$, and $dv_r/dr$ is the radial derivative of the line-of-sight component of the peculiar velocity. 

Below $z\sim25$ we have three broad periods reflected in the 21-cm signal (\citealt{PritchardReview}):
\begin{enumerate}
    \item WF coupling ($\lya$ pumping): The radiation from the first stars and galaxies begins to couple $T_{\rm S}$ to $T_{\rm K}$ via the WF effect. This drives the global signal ($\overline{\delta T_{\rm b}}$) into the absorption regime.
    
    \item X-ray heating: During the Epoch of Heating (EoH), the IGM is heated by X-rays. The $T_{\rm S}$, which is still tightly coupled to the $T_{\rm K}$, increases above $T_{\rm CMB}$ and the 21-cm GS shows an emission feature. The 21-cm signal also becomes insensitive to the spin temperature ($T_{\rm S}>>T_{\gamma}$ in equation (\ref{eq:T_b})). 
    
    \item Reionisation: As reionisation proceeds, the 21-cm signal goes to zero.
\end{enumerate}

All of these epochs are reflected in the 21-cm GS and 21-cm PS. X-rays can have a significant impact on the timing and extent of these periods, most notably the EoH.

As is evident from equation (\ref{eq:T_b}) the 21-cm signal depends on the ionisation, density, velocity, and spin temperature fields. We compute the $\hone$ 21-cm signal from the EoR by efficiently computing 3-D grids of 21-cm $T_{\rm S}$ and ionisation fields while the velocity and density fields are sourced from the N-body simulation.

Most studies in the literature including galaxy formation focus on the post-heating regime with the simplification that $T_\mathrm{S} >> T_{\gamma}$ (\citealt{Dragons5}; also see \citealt{GreigCMMC2} for a detailed analysis of the impact of this assumption). While likely valid during the late stages of the EoR, when the luminous sources have managed to couple the spin temperature to the kinetic temperature, for observations into the Dark Ages and the EoH this assumption breaks down. The main drivers of heating of the cosmic \hone are X-ray photons (\citealt{Furlanetto2006}, \citealt{McQuinn2012}). Large-scale simulations with low mass resolution are unable to simulate the effects of X-rays since the build-up of the stellar mass is delayed (as we demonstrate in section \ref{sec:sims_compare}). We use \meraxes combined with our augmented N-body simulations for calculations of the full brightness temperature field including contributions from heating, the spin temperature, recombinations, and peculiar velocities.
\subsection{Calibration}\label{sec:calibration}

\begin{figure}
\includegraphics[width=\columnwidth ]{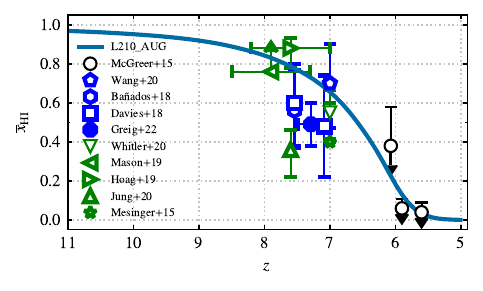}\vspace{-4mm}
\caption{\label{fig:xH_obs}Constraints on the reionisation history of the \genpsim simulation (blue curve). We use an evolving redshift-dependent escape fraction prescription for constraining the EoR history. The observational data are from analyses of dark pixels of $\lya$ \& $\lyb$ forests (\citealt{Mcgreer2015}), and $\lya$ damping wing absorptions (\citealt{Mesinger2015}
		\citealt{Banados2017}, 
        \citealt{Davies2018}, 
		\citealt{Mason2019}, 
		\citealt{Hoag2019},
		\citealt{Whitler2020}, 
		\citealt{Jung2020},
		\citealt{Wang2020},
		\citealt{Wold2022}, 
		\citealt{Greig2022}).
		}
\end{figure}

\begin{figure}
	\includegraphics[width=\columnwidth ]{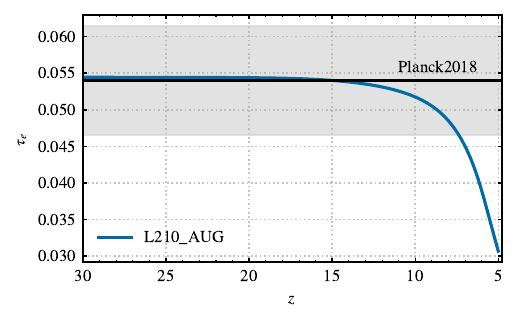}\vspace{-4mm}
		\caption{Figure shows the optical depth to CMB photons due to the free electrons. The blue curve is the integrated optical depth ($\tau_e$) computed from the fiducial \genpsim simulation. The black curve and the shaded region show the most recent measurement of $\tau_e$ from the \textit{Planck 2018} collaboration (\citealt{Planck2018}).
		}\label{fig:cmb_tau}
\end{figure}

Due to the numerous physical processes involved in galaxy formation and evolution, SAMs generally contain a large number of free parameters. In order to determine values for these parameters to ensure that a realistic galaxy population is produced, requires calibrating the model against a number of  existing observations. There are two different sets of calibrations involved in \meraxes -- one for the different galaxy formation parameters and the other for the reionisation calculations\footnote{Reionisation feedback affects low-mass galaxy formation but not properties constrained by observed LFs (eg. \citealt{Dragons3}).}. We calibrate the $\genpsim$ simulation by varying the following subset of input parameters of $\meraxes$: $\alpha_{\rm SF}$, $\sum_{\rm SF}$, $\eta_0$, $\epsilon_0$, $\gamma_{\rm DTG}$, $f_{\rm esc}$, \& $\alpha_{\rm esc}$ (see Table \ref{table:params} for details). We use the same parameter values for all of our simulations (listed in Table \ref{table:sims}).

\cite{Dragons19} calibrated the galaxy formation parameters of \meraxes against observed luminosity functions (LFs) and colour-magnitude relations at $z\sim 4 - 7$. In this work, we calibrate our simulations with respect to the LFs and the stellar mass functions (SMFs) in the $z\sim 5 - 8$ range. We find that, except for the $\gamma_{\mathrm{DTG}}$, the galaxy parameters from \cite{Dragons19} give a good fit to the data. Fig. \ref{fig:calibration}.b shows the dust attenuated luminosity functions for redshifts 8, 7, 6, \& 5 along with the observational data points (\citealt{Bouwens2015} and \citealt{Bouwens2021}). Fig. \ref{fig:calibration}.c shows the stellar mass functions for the corresponding redshifts with observations from \cite{Song2016} and \cite{Stefanon2021}.
We had to re-calibrate the $\gamma_{\mathrm{DTG}}$ (0.65 instead of 1.20; see Table 2 of \citealt{Dragons19}) parameter which governs the manner in which dust optical depth scales with the cold gas metallicity of the galaxy. The reason is that we extend our calibrations to brighter regions of the LFs than were available to \cite{Dragons19} because of their smaller simulation size. We summarise the parameters of \meraxes along with their values which have been used for calibration in Table \ref{table:params}.

\begin{table}
    \centering
    \begin{tabular}{c c c c }
    \hline
    Parameter & Value & Description\\ \hline
    \vspace{-2mm}\\

    $\alpha_{\rm SF}$ & 0.10 & Star formation efficiency \\

    $\Sigma_{\rm SF}$ & 0.01 & Critical mass normalisation \\

    $\eta_0$ & 7.0 & Mass loading normalisation \\
    
    $\epsilon_0$ & 1.5 & Supernova energy coupling normalisation\\
    
    $\gamma_{\mathrm{DTG}}$ & 0.65 & Galaxy metallicity scaling of optical depth \\
    \vspace{-2mm}\\
    \hline
    \vspace{-2mm}\\

    $f_{\mathrm{esc}}^o$ & 0.14 & Escape fraction normalisation \\
    $\alpha_{\mathrm{esc}}$ & 0.2 & Escape fraction redshift scaling \\
    \hline    
    \end{tabular}
    \caption{The fiducial input parameters and their values used for the simulations listed in Table \ref{table:sims}. The first set of these ( $\alpha_{\rm SF}$, $\sum_{\rm SF}$, $\eta_0$, $\epsilon_0$, \& $\gamma_{\rm DTG}$) are calibrated to the observed LFs and SMFs and control the galaxy properties of $\meraxes$, while $f_{\rm esc}$ \& $\alpha_{\rm esc}$ is calibrated with respect to the reionisation constraints. See section \ref{sec:calibration} for more details.}
    \label{table:params}
\end{table}

The second set of calibrations is for the reionisation calculations. The photon budget is influenced by the escape fraction ($f_{\rm esc}$) of the galaxies which sets the fraction of photons that are able to survive the absorption by dust and neutral gas in and around the galaxies and escape into the IGM. The high-$z$ escape fraction is one of the least constrained parameters in the literature. In this work, we use a prescription that is skewed towards the high redshifts as the shallower potential of the small galaxies at high-$z$ results in more photons escaping their hosts. Additionally, the impact of the Monte Carlo haloes is more relevant at high-$z$ and this implementation helps to bring out the importance of these galaxies. In this work use an $f_{\rm esc}$ that evolves with redshift $z$ as:
\begin{equation}
f_{\rm esc} = f_{\rm esc}^{o} \bigg (\dfrac{1+z}{6}\bigg )^{\alpha_{\rm esc}},
\end{equation}
where $f_{\rm esc}^o$ is the escape fraction normalisation and $\alpha_{\rm esc}$ sets the escape fraction redshift scaling. We tune these parameters such that our reionisation history matches the measured constraints on the IGM neutral fractions (fig. \ref{fig:xH_obs}) and the integrated optical depth of CMB photons ($\tau_e$) from scattering off free electrons (Fig. \ref{fig:cmb_tau}; \citealt{Planck2018}).

\section{Reionisation Predictions}\label{sec:sims_compare}

In this section, we demonstrate the full reionisation model from our \genpsim simulation. In particular, we focus on comparing the impact of the missing  low-mass haloes from the \gensim simulation to illustrate the importance of mass resolution.\footnote{The reader is cautioned that in the current implementation of \meraxes we are only forming and evolving atomically cooled galaxies and are thus missing the possible contribution from smaller galaxies in molecularly cooled haloes (so-called mini-haloes) which are likely to contain PopIII stars. These mini-haloes can contribute to the build-up of the background radiation fields and will have an impact on reionisation (see for example \citealt{Yuxiang2020}, \citealt{Yuxiang2021}, \citealt{EmanueleMaria_pop3_1}). Thus discussions on the appearance of features in this work will also be delayed relative to simulations which also include mini-haloes.}

Since we are missing smaller mass haloes below the mass resolution in \gensim we delay the build-up of cosmic stellar mass within galaxies. Thus, there is also a delay in any physical property that is dependent on the total stellar mass (such as ionisations and radiation backgrounds). This will result in the X-ray background forming too late, and correspondingly the IGM cools for longer before heating. However, the \genpsim simulation includes a realistic galaxy population capable of producing the whole X-ray background. We, therefore, present here the first large-scale ($>200 \oneh$ Mpc) simulation of the thermal and ionisation history of the cosmic $\hone$ incorporating realistic galaxy formation and evolution physics at masses down to the atomic cooling limit.

As the first application of this simulation, we explore the evolution of the volume-averaged neutral fraction $\xH$, the 21-cm GS and the 21-cm PS in this section. 

\subsection{EoR history and ionisation morphology}
\begin{figure}
		\includegraphics[width=\columnwidth ]{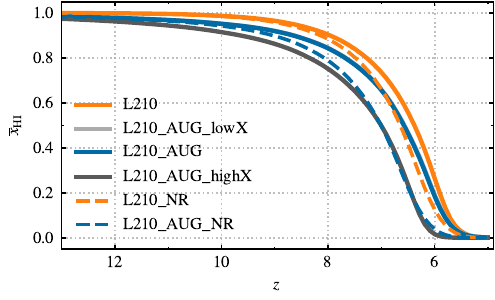}\vspace{-4mm}
		\caption{
		\label{fig:eor_hist}Global reionisation history of the \genpsim (solid blue) and \gensim (solid orange) simulations. \genpsim starts reionising earlier and also has a much more extended reionisation phase. We also vary the galaxy X-ray luminosity in our model (\pgenpsim \& \mgenpsim in dark and light grey respectively). The \mgenpsim is almost identical to \genpsim and thus the curves overlap. The two dotted curves are the same as \genpsim \& 
	\gensim except that we do not include inhomogeneous recombinations.}
\end{figure}
Fig. \ref{fig:eor_hist} shows the resultant EoR histories. As shown in Fig. \ref{fig:xH_obs}, by construction these resultant reionisation histories are consistent with all existing limits and constraints on the IGM neutral fraction during the reionisation epoch. The \genpsim box starts to reionise much earlier than \gensim owing to the introduction of low-mass galaxies found only in Monte-Carlo haloes. However, the fiducial \genpsim and \gensim simulations both finish reionisation at approximately the same redshift. There are two main reasons for this. Firstly, towards the end of the EoR, reionisation is primarily maintained by larger mass haloes (which are accurately simulated across both simulations) while the lower mass galaxies are more relevant at earlier times. Secondly, the impact of inhomogeneous recombination in the two simulations is different. Since small galaxies, which initiate reionisation in $\genpsim$ are short-lived, the cosmic gas recombines until sufficiently big galaxies have had time to form and produce enough ionising photons to complete reionisation. In order to check the role of recombination we, therefore, ran two additional simulations \nrgenpsim and \nrgensim (shown with dotted lines) where we have turned off inhomogeneous recombinations (i.e. setting $\overline{n}_{\rm rec} = 0$ in equation \ref{eq:excursion}). Turning off inhomogeneous recombination results in \nrgenpsim reionising much earlier than \nrgensim as expected. 

\begin{figure*}
		\includegraphics[width=\textwidth]{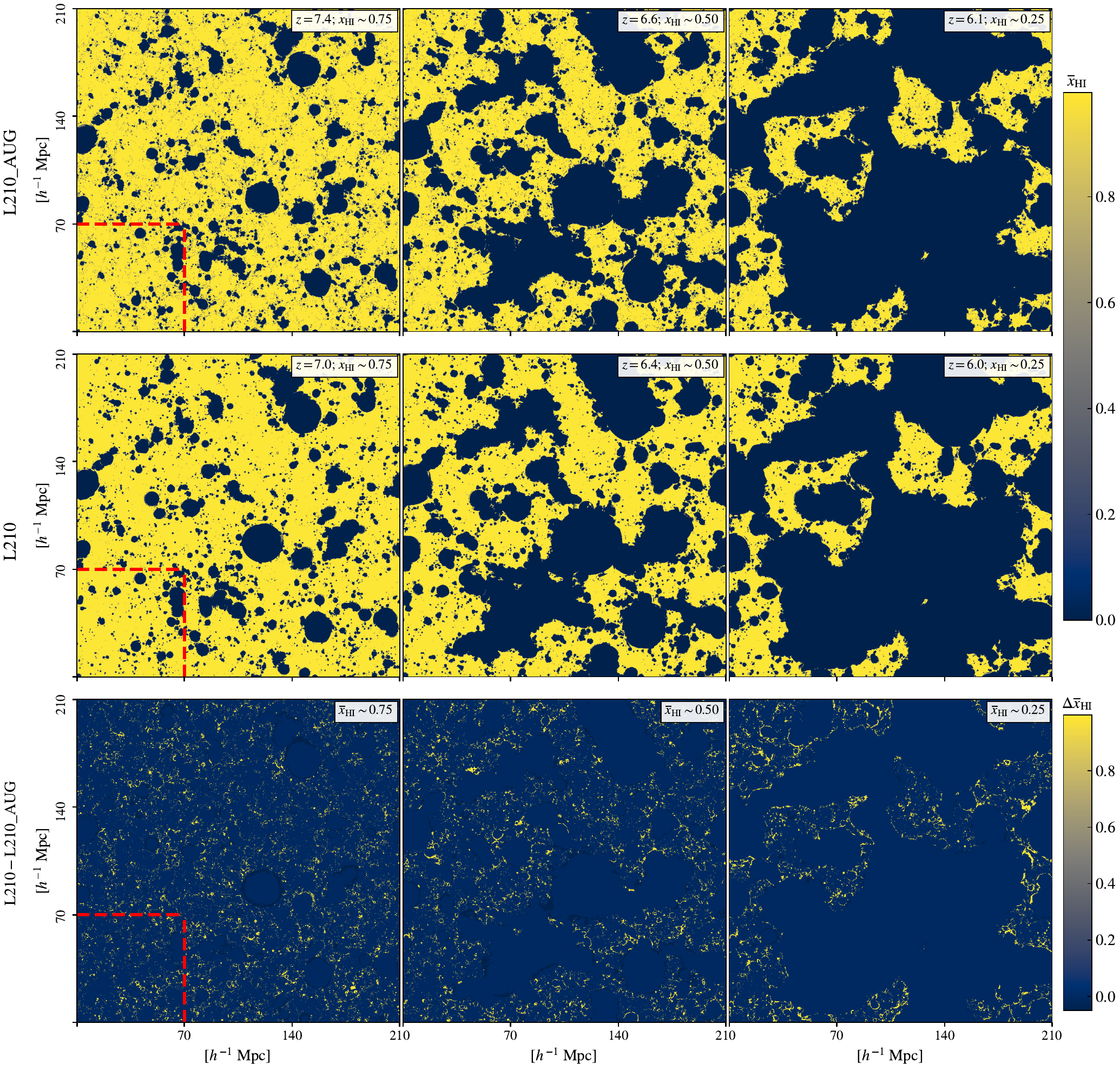}\vspace{-4mm}
		\caption{\label{fig:xH_field}2D slices of the $\xH$ grid from the \genpsim (upper row) and \gensim (bottom row) simulations. From left to right, the columns correspond to fixed neutral fractions,  $\xH \sim$ 0.75, 0.5, 0.25. For a particular $\xH$, \genpsim has a higher redshift compared to $\gensim$ because of the higher number of ionising photons. Yellow represents neutral hydrogen $(\hone)$, and blue regions are ionised hydrogen $(\hii)$ bubbles. Each slice is $210$ $\oneh$ Mpc on a side and $\sim 0.2$ $\oneh$ Mpc thick. The last row shows the difference between the ionisation fields of the two simulations ($\Delta \xH = \overline{x}_{\hone,\gensim} - \overline{x}_{\hone,\genpsim}$).
		We have used a colour gradient that is weighted towards the small scale structures to highlight the small $\hii$ regions that are due to the Monte-Carlo haloes. The red dashed regions in the first column show the size of our 70 $\oneh$ Mpc side-length subvolumes (see section \ref{sec:cv} for further details).
		}
\end{figure*}

Fig. \ref{fig:xH_field} shows slices from the ionisation fields of \genpsim (top row) and \gensim (middle row) for $\xH \sim $ 0.75, 0.50, \& 0.25. Each slice is $210 \oneh$ Mpc on a side and $\sim0.2\oneh$ Mpc thick. In order to emphasise the effects of the introduction of the small haloes we compare the simulations at the same volume-averaged neutral fractions ($\xH$). At any given $\xH$ there will be more small haloes in \genpsim compared to  $\gensim$. One of the main impacts of the smaller haloes is to force reionisation to begin earlier as discussed above. Thus, on average the large ionised regions in \genpsim are smaller as they are driven by lower stellar masses relative to the \gensim simulation as indicated by the earlier redshift for a fixed neutral fraction. The third row shows the difference between the two simulations with the colour gradient chosen to highlight the contribution from the extra small mass haloes. In the ``Difference" (bottom row) we have subtracted the ionisation fields of \genpsim from \gensim to clearly bring out the impact of the augmentation.

\subsection{21-cm statistics}

Fig. \ref{fig:global_xray} shows the 21-cm GS for $\genpsim$ and $\gensim$. We find that the \genpsim simulation has a similar (but broader) absorption feature, though occurring earlier in redshift, relative to $\gensim$. This highlights the importance of introducing the low-mass haloes beyond the resolution limit of $\gensim$. By including these in $\genpsim$, the $\lya$ and X-ray background builds up at earlier times due to the additional low-mass haloes. The $\lya$ background couples the spin temperature to the gas kinetic temperature $T_{\rm K}$, which is much lower than the CMB temperature $T_{\rm CMB}$, resulting in the broader and earlier absorption. We also point out that the gradient of the absorption feature in the 21-cm GS is larger in \gensim as compared to $\genpsim$. The delayed but sudden formation of sources in $\gensim$, relative to $\genpsim$, results in a comparatively rapid build-up of the stellar mass and consequently the radiation backgrounds.

Fig. \ref{fig:21ps_xray} compares the evolution of the spherically-averaged 3-D 21-cm PS at fixed spatial scales ($k\sim 0.1$ $\rm Mpc^{-1}$ and $k\sim 1$ $\rm Mpc^{-1}$) for the \gensim and \genpsim simulations. For both of these scales, the evolution of the power for \gensim is qualitatively similar to $\genpsim$. However, the delayed formation of stellar mass in \gensim results in there being considerable differences between the timing of the peaks. The \lya-coupling peak in \gensim is delayed relative to \genpsim by $\Delta z \sim 3$. Below $z\sim7$, when the X-rays have already initiated the EoH and EoR is well on its way, the power in both \genpsim  and \gensim becomes similar.

Even though the large-scale ($k\sim 0.1$ Mpc$^{-1}$) 21-cm power is expected to have three peaks corresponding to the \lya-pumping, X-ray heating, and reionisation epochs (see \citealt{Pritchard2007}, \citealt{MesingerXRay}) we observe only two peaks in our simulations. The EoH peak, expected at $z \sim 12$ for \genpsim (corresponding to the minima of the global signal; see Fig. \ref{fig:global_xray}) is masked by the $\lya$ peak. They have merged together into one broad peak owing to the timing and build-up of the backgrounds during these two epochs\footnote{Another contributing factor is the slow build-up of the backgrounds owing to the time-scales of ionising photons (as evidenced by the different gradients of the global signals).}.

On the other hand, the redshift evolution of the 21-cm power on small-scales ($k\sim 1$ Mpc$^{-1}$) is characterised by two peaks corresponding to a combined \lya-pumping and EoH, and an EoR peak (\citealt{Yuxiang2020}). Typically, on small scales, the impact of the EoH is harder to disentangle as it primarily impacts larger scales due to the larger mean free path of X-ray photons.

\subsection{Effects of varying the Galaxy X-Ray Luminosity in the early Universe}

The large volume of our simulations enables the exploration of the effects of X-rays on the EoR morphology with a full source population. Here, we only vary $L_{X<\text{2 keV}}/\mathrm{SFR}$, keeping $E_0$  and $\alpha_X$ fixed. In addition to our fiducial value of $3.16 \times 10^{40}$ erg $\rm s^{-1} \Msun^{-1} \rm{yr}$, we also consider $3.16 \times 10^{38}$ and $3.16 \times 10^{42}$ erg $\rm s^{-1} \Msun^{-1} \rm{yr}$ which we call \mgenpsim and \pgenpsim simulations respectively (see table \ref{table:sims}). This enables us to encompass the plausible range of contribution of the X-rays in the early Universe (\citealt{Fialkov2017}, \citealt{GreigCMMC2}). Fig. \ref{fig:lightcones} shows the light cone evolution (of $\delta T_b$) for these simulations. We note that \citealt{HERA_limits} has recently ruled out a number of "cold reionisation" models  corresponding to low X-ray luminosity. In light of this, our \mgenpsim model is a very unlikely scenario. However, our aim in this work is to develop an intuition for the impact of X-rays in the early Universe from a galaxy SAM and reionisation simulation.

In the next subsections, we compare the impact of the X-ray photons in the early Universe relative to our fiducial model.

\subsubsection{EoR history}

The ionisation photon budget is dominated by UV photons, with the X-rays contributing at most $10-15$ per cent in the most extreme models (\citealt{MesingerXRay}). Fig. \ref{fig:eor_hist} shows the reionisation histories from all three of the augmented simulations. We find that, in agreement with studies in the literature (\citealt{MesingerXRay}), though the role of X-rays in ionising the $\hone$ is much less than that of the UVB, they can hasten reionisation (see \genpsim vs \pgenpsim in Fig. \ref{fig:eor_hist}). 

\subsubsection{21-cm Light Cone}\label{sec:lightcone}
\begin{figure*}
		\includegraphics[width=\textwidth]{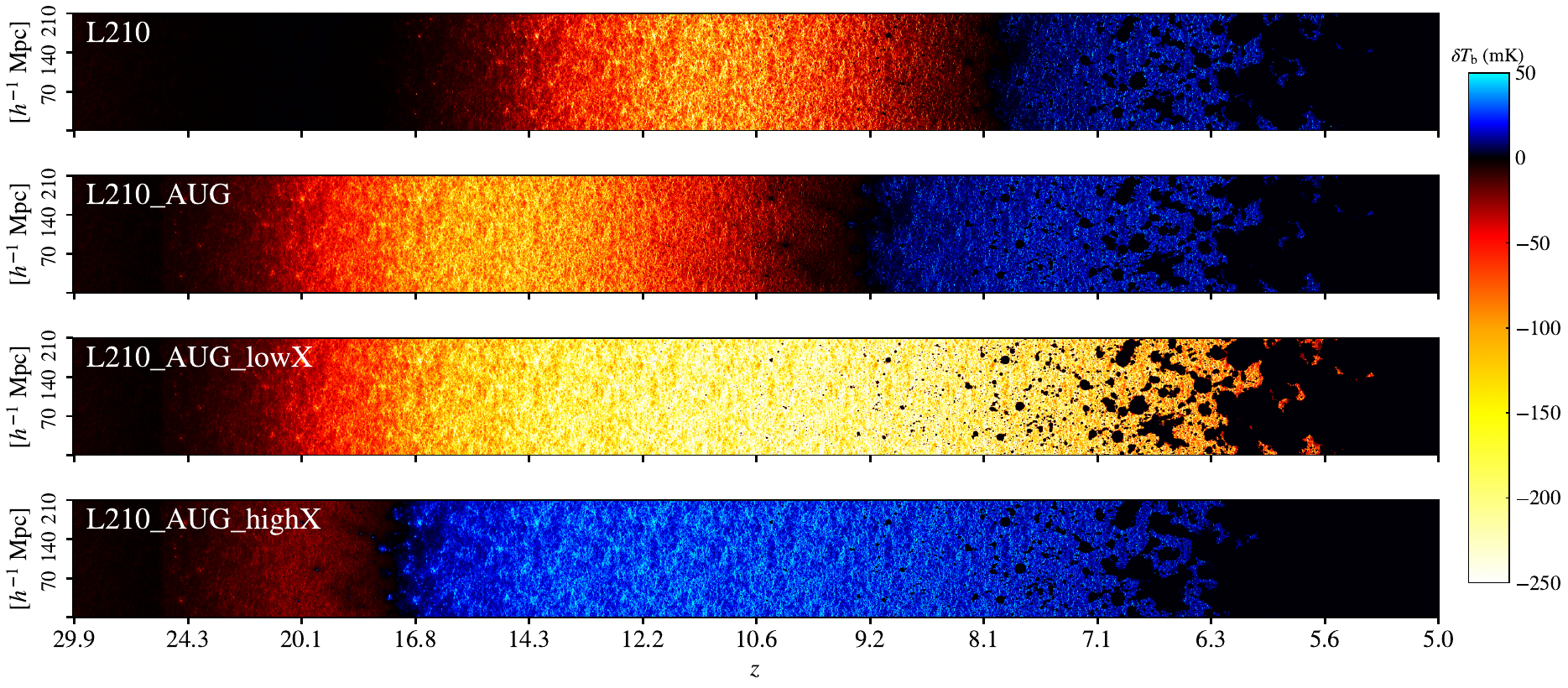}\vspace{-4mm}
		\caption{
		\label{fig:lightcones}The light cone evolution of the 21-cm brightness temperature ($\delta T_b$) from our simulations. $\gensim$ is characterised by the delayed but rapid evolution of $\delta T_{\rm b}$ because of its lower resolution. We also point out $\mgenpsim$, characterised by low galaxy X-ray luminosity, in which the cosmic $\hone$ remains cold and never goes into emission.}
\end{figure*}
In Fig. \ref{fig:lightcones} we show the 21-cm brightness temperature ($\delta T_{\rm b}$) light cone slices from our simulations. The 21-cm signal, being a line transition, evolves along the line-of-sight and light cones provide a realistic representation of the evolution of such cosmic signals. We stitch together the $\delta T_{\rm b}$ grids from our coeval simulation boxes to generate the light cone by linearly interpolating them in cosmic time between snapshots. The delayed but rapid evolution in the case of $\gensim$, compared to the rest of the simulations, underscores the impact of the mass resolution on the 21-cm signal. For $\mgenpsim$, the signal remains in absorption across our full redshift range whereas for $\pgenpsim$ it is mostly in emission.

\subsubsection{21-cm Global Signal}

\begin{figure}
		\includegraphics[width=\columnwidth]{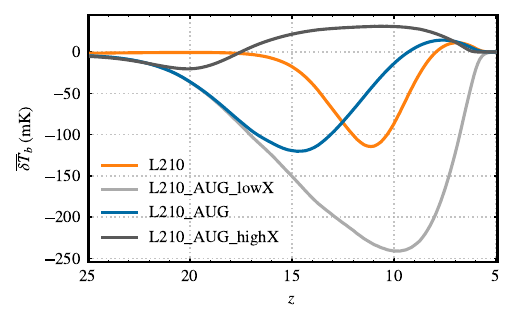}\vspace{-4mm}
		\caption{
		\label{fig:global_xray}Figure shows the effect of X-rays on the 21-cm GS from the cosmic dawn and EoR. As expected, more X-rays (dark grey) cause the signal to be observed in emission earlier whereas a lack of X-rays (light grey) causes a deeper absorption feature.
		 }
\end{figure}

Fig. \ref{fig:global_xray} shows the 21-cm GS from all four simulations. As shown in section \ref{sec:lightcone}, the main physical impact of X-ray photons is to heat the cosmic gas. With respect to our fiducial $\genpsim$ simulation, \mgenpsim simulation has less X-ray photons resulting in the cosmic gas being colder, and hence a stronger absorption dip. All of our simulations except $\mgenpsim$ go into emission (also evident from Fig. \ref{fig:lightcones}). $\pgenpsim$, being characterised by significantly more X-ray photons than the other simulations, has a hot IGM resulting in the signal going into emission at $z \lesssim 17$ as well as the merging of the \lya-pumping epoch with the EoH.

\subsubsection{21-cm Power Spectra}
\begin{figure*}
	\begin{minipage}{\textwidth}
		\centering
		\includegraphics[width=\textwidth]{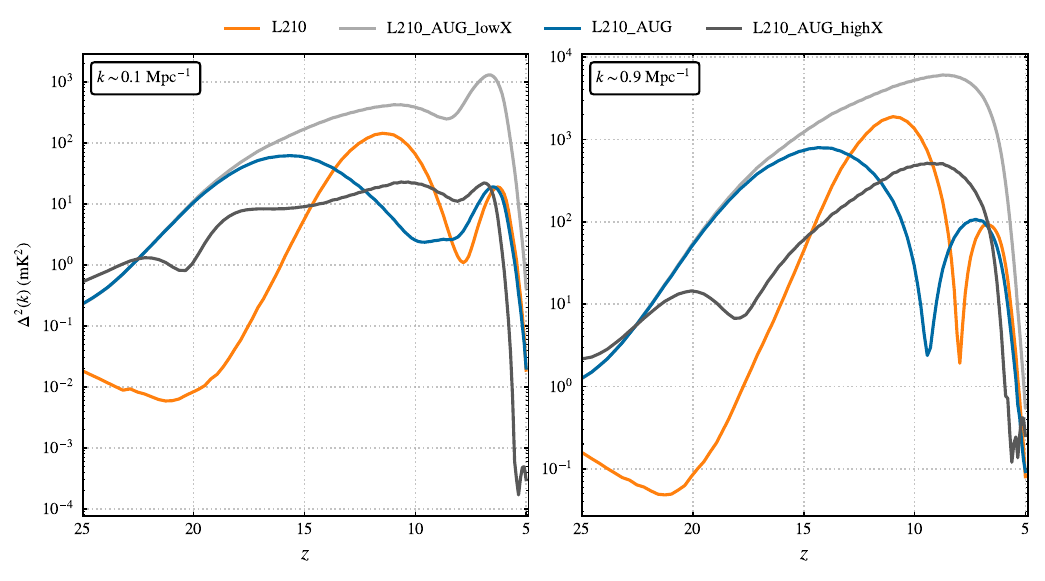}\vspace{-4mm}
		\caption{
		\label{fig:21ps_xray} We show the 21-cm power spectra for the simulations at two scales ($k\sim 0.1$ $\rm Mpc^{-1}$  on left and $k\sim 1$ $\rm Mpc^{-1}$ on the right).}
	\end{minipage}
\end{figure*}
In Fig. \ref{fig:21ps_xray} we compare the evolution of the 21-cm PS at $k\sim 0.1$ $\rm Mpc^{-1}$ and $k\sim 1$ $\rm Mpc^{-1}$. Like the 21-cm GS, the shape and amplitude of the 21-cm PS are strongly affected by the X-ray luminosity. An accurate measurement of the 21-cm PS will thus have great constraining power on the properties of the X-ray sources in the early Universe (see, for example, the recent constraints from \citealt{HERA_limits}).

At large scales (left panel of Fig. \ref{fig:21ps_xray}) we observe the expected features, though there is considerable variation in the timing and duration among the simulations. We note that the \mgenpsim simulation has the highest power for most epochs (with peak power during the EoR at $z\sim7$) owing to the large temperature contrasts due to the cold IGM. The inefficient heating because of the low X-ray luminosity has also resulted in the EoH and EoR peaks merging together. \genpsim simulation shows the 3 expected peaks with the power peaking at $z\sim 16$ corresponding to the absorption in the 21-cm GS; there is thus more power during the $\lya$-pumping epoch. \pgenpsim is characterised by less power during all epochs. Though the general features are similar to the \genpsim simulation, the amplitude and timing are different due to the reduction in the amplitude of the IGM temperature contrast. The \pgenpsim simulation is characterised by much smaller temperature fluctuations than the other two simulations. Thus, during the EoH this simulation has the smallest amplitude.

The right panel shows the power on small scales. The power on this scale exhibits the expected behaviour. Interestingly, \pgenpsim has clearly differentiated \lya-pumping and EoH peaks. The EoH peak in this case has merged with the EoR peak, owing to the extended EoH because of the large X-ray luminosity.

\section{Cosmic Variance in EoR Statistics}\label{sec:cv}
Measurement of any statistical signal from a finite volume of the Universe introduces an inherent uncertainty in its variance since we are only sampling one realisation of the underlying statistical ensemble. This is termed the \textit{cosmic variance}. In this section, as an application of our large volume simulations, we explore the cosmic variance of the 21-cm signal.

To explore the cosmic variance we divide each of the augmented simulations into 27 equal sub-volumes each of side $70 \oneh$ Mpc. Each sub-volume is larger than the typical largest ionised regions even during the late stages of reionisation. The $70 \oneh$ Mpc sub-volumes are also comparable to most state-of-the-art radiation hydrodynamical simulations in the literature (\citealt{CROC3}, \citealt{BLUETIDES},  \citealt{SpringelTNG}, \citealt{CODA},  \citealt{THESAN}) and also to the largest simulation volume on which $\meraxes$ has been run (\citealt{Dragons19}) as part of the DRAGONS project. Our ensemble, therefore, provides an estimate of the cosmic variance in these simulations.

\subsection{EoR History and 21-cm Global Signal}
\begin{figure}
		\includegraphics[width=\columnwidth]{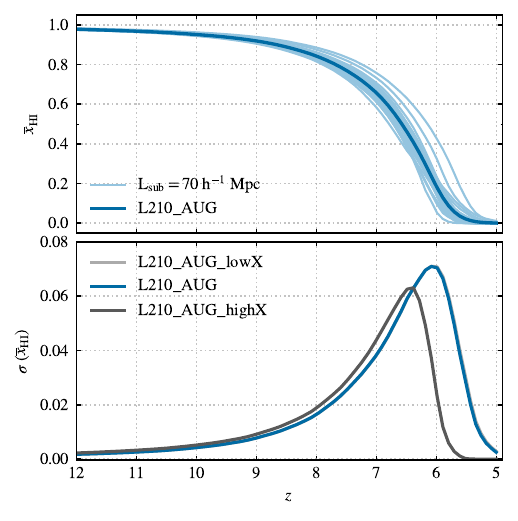}\vspace{-4mm}
		\caption{\label{fig:xH_subvols} The {\color{blue}blue} curve shows the EoR history of the \genpsim simulation and the lighter shades show the EoR history in the 27 subvolumes which the \genpsim has been divided into. We find a spread in the EoR history with almost $\Delta z \sim 0.8$ around $\xH \sim 0.5$. The bottom panel shows the standard deviation of the \xH ($\sigma(\xH)$) among the subvolumes and we show this for all the augmented simulations. The \mgenpsim (light grey) curve is identical to the \genpsim curve and lies behind it.}
\end{figure}
\begin{figure}
		\includegraphics[width=\columnwidth ]{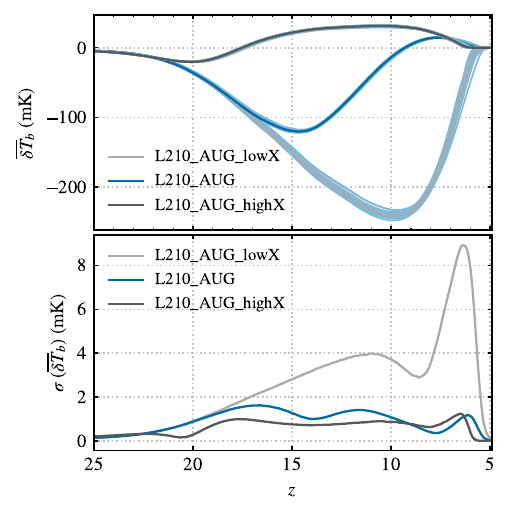}\vspace{-4mm}
		\caption{
		\label{fig:global_std}The evolution of the 21-cm GS among the subvolumes (light blue) for all three of the augmented simulations.The bottom panel shows the standard deviation of the global signal among the 27 subvolumes. 
		}
\end{figure}
Fig. \ref{fig:xH_subvols} shows the spread in the reionisation histories for the different subvolumes (in light-blue) relative to \genpsim (in blue). We find that the range in redshift for reionisation histories among sub-volumes at $\xH\sim0.5$ is $\Delta z \sim 0.8$. The bottom panel shows the standard deviation of $\xH$ ($\sigma(\xH)$) among the subvolumes. We see the same trend among the simulations except that features in the \pgenpsim simulation occur earlier relative to the \genpsim and \mgenpsim simulations (which are almost identical). 

In Fig. \ref{fig:global_std} we do a similar analysis for the 21-cm GS with the top panel showing the signal from the subvolumes and the bottom panel showing the standard deviation.
Comparing the bottom panels of Fig. \ref{fig:xH_subvols} \& Fig. \ref{fig:global_std} we see that the fractional change in $\overline{\delta T}_{\rm b}$ is higher than in $\xH$. During the EoH, the scatter in $\delta T_{\rm b}$ is driven by variations in $T_{\rm S}$ while during reionisation the scatter in $\xH$ dominates.

\subsection{21-cm Power Spectra}

\begin{figure*}
		\includegraphics[width=\textwidth]{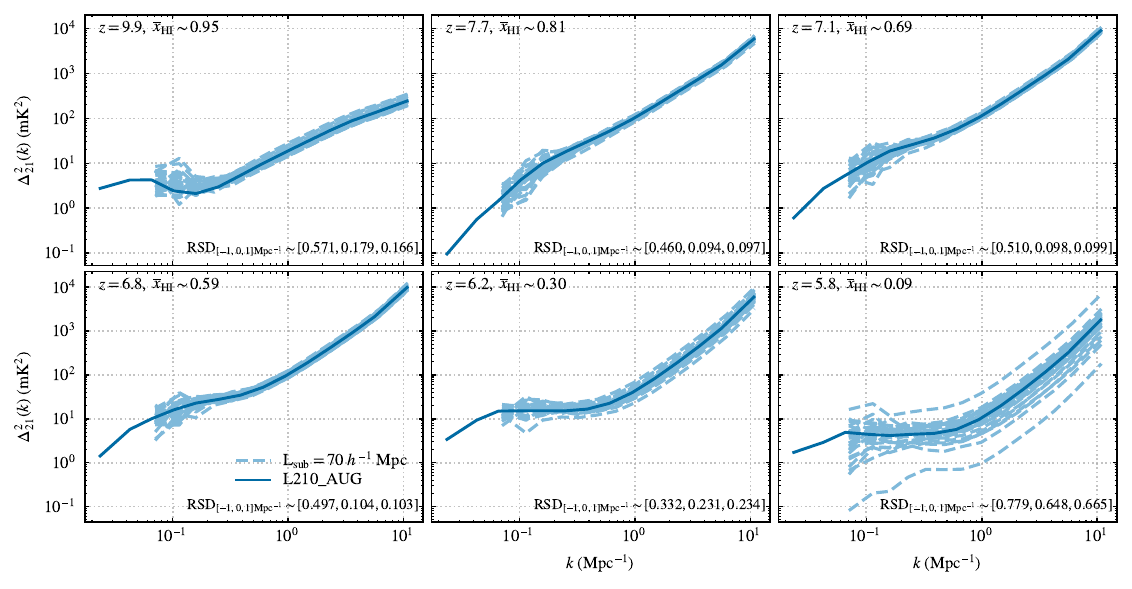}\vspace{-4mm}
		\caption{
		\label{fig:ps_subvol}The fiducial augmented simulation \genpsim has been subdivided into 27 equal subvolumes. Shown (dashed light blue) here are the 21-cm power spectra from these subvolumes. The power spectra from the whole volume are also shown (solid blue). The subplots correspond to $\xH \sim 0.95, 0.8, 0.70, 0.50, 0.30, 0.10$. We also show the relative standard deviation ($\rm{RSD}=\frac{\text{standard deviation}}{\rm{mean}}$) of the powers at $k= 10^{-1}, 10^{0}$ \& $10^{1}$ $\rm{Mpc}^{-1}$ spatial scales from the subvolumes.}
\end{figure*}

Fig. \ref{fig:ps_subvol} shows the 21-cm PS from the \genpsim simulation (in blue) at $\xH\sim$ 0.95, 0.8, 0.70, 0.60, 0.30, and 0.10. The 21-cm PS from the 27 subvolumes (in light blue) are also shown. The scatter in the 21-cm PS increases for decreasing $k$-value (towards large scales) and decreasing redshift (as reionisation progresses). The spread in power for large $k$-values is larger than the spread at small $k$-values for all redshifts. This is due to sample variance since there are fewer modes at these large scales in the volumes to average over. At low redshifts, most of the 21-cm emission comes from sparse, isolated neutral patches leading to considerable scatter in the 21-cm power. 

\begin{figure*}
		\includegraphics[width=\textwidth]{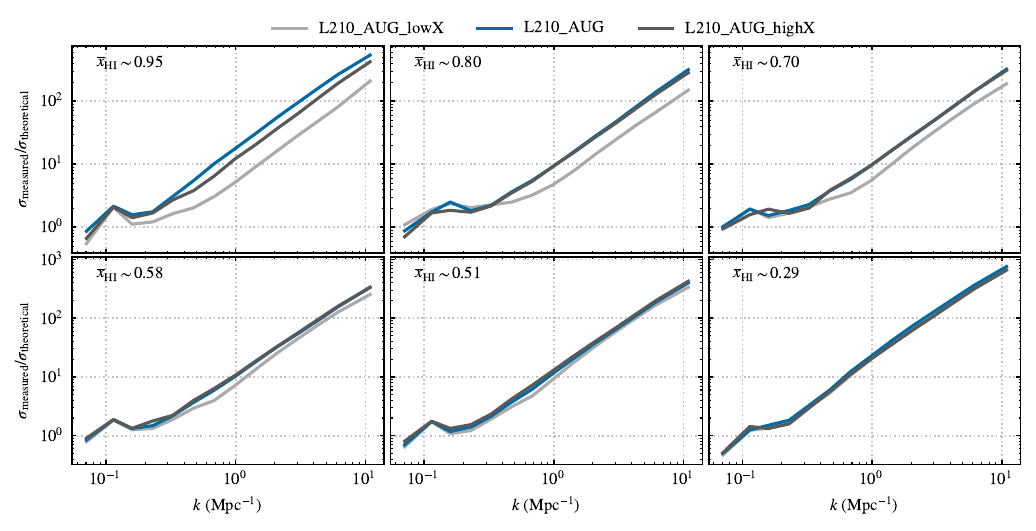}\vspace{-4mm}
		\caption{Figure shows the ratio of measured to theoretical errors. We compute the standard deviation of the power among the 27 subvolumes as a function of $k$, $\sigma_{\rm measured}$. We compare this with the $\sigma_{\rm theoretical}$ where we use the mean of the powers among the subvolumes as the $\overline{P}(k_i)$ in equation (\ref{eq:deltap}).}
		\label{fig:21ps_xray_err_ratio} 
\end{figure*}
The power spectrum quantifies the variance in amplitudes of a random field on different scales. A purely Gaussian-random field is fully specified by its power spectrum (\citealt{PeeblesLSS}). The cosmic variance of the power spectrum in this case should simply be the Poisson sampling error which depends only on the number of modes in each spherical shell in $k$-space. However, higher-order statistics are required to capture the information for non-Gaussian fields.

The 21-cm field is non-Gaussian, especially on small scales and during the final stages of the EoR. Initially, the 21-cm emission traces the underlying matter-density field which is Gaussian on large scales where the evolution is governed by linear theory. However, once the complex 3D morphology of the radiation fields (e.g. ionisation, X-ray or Lyman-alpha) begins to impact the 21-cm signal, the statistics will deviate from Gaussianity (\citealt{StuartReview}). Hence the cosmic variance of the 21-cm power spectrum will be larger than the Poisson sampling error. Here, we explore the impact of  non-Gaussianity on the cosmic variance uncertainty of the 21-cm PS. 

\cite{Mondal2016} showed that non-Gaussianity has an appreciable impact on the error-covariance of the power spectra. The full error-covariance matrix of the 21-cm PS is given by
\begin{equation}\label{eq:c_ij}
      \pmb{C}_{ij} = \dfrac{1}{V}\Bigg[\Bigg( \dfrac{(2\pi)^2[\overline{P}(k_i)]^2}{k_i^2\Delta k_i}\Bigg) \delta_{ij}+ \overline{T}(k_i, k_j)\Bigg],
\end{equation}
where $V$ is the simulation volume, $k_i$ is the average spatial frequency in the $i$th bin, $\Delta k_i$ is the bin-width of the $i$th bin,  $\overline{P}(k_i)$ is the power spectrum averaged over the $i$th bin, and $\overline{T}(k_i, k_j)$ is the average trispectrum. This trispectrum component arises from the non-Gaussianity of the 21-cm signal.

Generally, studies in the literature make the simplifying assumption that the 21-cm field is Gaussian and ignore the second term in equation (\ref{eq:c_ij}) giving

\begin{equation}\label{eq:deltap}
\begin{split}
    \delta P(k_i) & = \sqrt{\pmb{C}_{ii}} \\
    & = \sqrt{\dfrac{(2\pi)^2[\overline{P}(k_{i})]^{2}}{V k_{i}^{2} \Delta k_{i}}}.
\end{split}
\end{equation}
Hence, any deviation from equation (\ref{eq:deltap}) measured from our 27 sub-volumes must occur as a result of the non-Gaussianity of the 21-cm signal.

From each of the 27 equal sub-volumes, we compute the spherically averaged power spectrum and show in Fig. \ref{fig:21ps_xray_err_ratio} the ratio of the measured cosmic variance from the sub-volumes ($\sigma_{\rm measured}$) to that expected theoretically from equation (\ref{eq:deltap}) ($\sigma_{\rm theoretical}$). Specifically, $\sigma_{\rm measured}$ is computed as the standard deviation of the power among the subvolumes as a function of $k$, while $\sigma_{\rm theoretical}$ is computed using equation (\ref{eq:deltap}) where the $\overline{\rm P}(k_i)$ is the mean 21-cm PS from the subvolumes. For a Gaussian field, we expect the ratio $\sigma_{\rm measured}/\sigma_{\rm theoretical}$ to be unity. We provide this ratio for $\xH\sim0.95$, 0.75, 0.50, and 0.25 in each of our three augmented simulations.

We find similar features and trends among all of our simulations. The ratio\footnote{At the same time we caution that it is possible we over-predict the cosmic variance due to the smaller size of our sub-volumes. Our sub-volumes are $70 \oneh$ Mpc, which are slightly smaller than expected for convergence of the statistics. This may also explain why our ratio $\sigma_{\rm measured}/\sigma_{\rm theoretical}$ sits above unity for the largest scales (i.e. where it is expected to be Gaussian).} increases from small to large $k$-values implying that the large-scales are more Gaussian in nature compared to the small scales. 

Our results agree qualitatively with \cite{Mondal2015} and \cite{Mondal2016} who show that the non-Gaussianity of the 21-cm field grows with the progress of reionisation. During the early stages of the EoR ($\xH\sim0.80$ case), we find that the contribution of non-Gaussianity to the variance of the 21-cm PS is comparable to the Gaussian term for $0.1\lesssim k \lesssim 0.4$ $\rm Mpc^{-1}$ (where the ratio $\sigma_{\rm measured}/\sigma_{\rm theoretical} \sim \mathcal{O}(1)$) while for $k\gtrsim 2$ $\rm{Mpc}^{-1}$ the ratio is $>10$. As EoR progresses (see $\xH\sim0.30$ subplot), this ratio becomes $2-4$ for $0.1\lesssim k \lesssim 0.4$ $\rm Mpc^{-1}$ and up to $\sim 100$ for $k\gtrsim 2$ $\rm{Mpc}^{-1}$. At the same time, we find that the transition to non-Gaussianity in our model appears to occur earlier than in \cite{Mondal2016}. Likely this is a result of the detailed physics prescription of our model, making a direct comparison hard particularly at high redshifts since our simulations include spin temperature fluctuations which likely add to the non-Gaussianity in the 21-cm signal.

Our results show that when estimated by assuming that the 21-cm field is Gaussian (i.e. using equation \ref{eq:deltap}), towards the end of EoR, the cosmic variance within $\sim100$ Mpc boxes is underestimated by a factor of $\sim2$ within $k\sim 0.1-0.5$ Mpc$^{-1}$ scales which are the main focus of the current and upcoming telescopes observing 21-cm fluctuations.

\section{Conclusion}\label{sec:conclusion}
In this paper, we have introduced an updated version of the \meraxes semi-analytic model, which for the first time includes heating from X-rays and thermal evolution of gas in the IGM. In order to have sufficient volume for calculating the effect of X-rays during reionisation we utilise a new, large-volume N-body simulation with side-length $L = 210 \oneh$ Mpc and $4320^3$ particles ($\gensim$). To resolve all atomically cooled haloes out to $z=20$ necessary for studying galaxy formation (of $\sim2\times10^7\oneh \Msun$) we performed Monte-Carlo augmentation of this simulation using \dforest ($\genpsim$). This achieves an effective N-body particle number of $\sim10,000^3$. $\genpsim$ is a unique dataset for exploring galaxy formation physics and its impact on the timing and morphology of the EoR. Coupling $\meraxes$ to this augmented simulation enables the exploration of the different galaxy formation parameters on the 21-cm signal. We found that the inclusion of these Monte-Carlo haloes has a significant impact on the build-up of stellar mass in our simulations and consequently on reionisation, which commences earlier and is more gradual. We also find that \lya-coupling and X-ray heating, and hence the end of the 21-cm global minima occur earlier in the higher resolution simulation. In addition, we find that the timing and duration of the peaks of the 21-cm power spectrum (PS) are different in the augmented higher-resolution simulation. These results underscore the need for both large volume and sufficient mass resolution for simulations exploring the EoR.

The large volume of our simulation and the implementation of thermal and spin temperature evolution in \meraxes enables exploration of the impact of X-ray luminosity on heating the \hone gas. In agreement with semi-numerical studies (\citealt{MesingerXRay}, \citealt{GreigCMMC2}) we show that while their impact on the reionisation history is minimal X-rays can have an appreciable impact on both the 21-cm GS and on the 21 PS. Observations of the 21-cm PS will thus provide constraints on the X-ray properties of the sources in the early Universe.

Taking advantage of the large volume of our simulation, we explore the scatter in the reionisation history and the 21-cm global signal within 27 sub-volumes of side 70 $\oneh$ Mpc, which are each comparable to our previous simulations and state-of-the-art radiation-hydrodynamical simulations in the literature. We compare the standard deviation in the 21-cm PS amongst these sub-volumes to the Gaussian expectation for the variance of a random field. As previously described in \cite{Mondal2016}, we find that the non-Gaussianity of the signal contributes significantly to the variance of the 21-cm PS on all scales and increases towards the small scales. However, this work is the first study of the error-variance of the 21-cm PS at high redshifts in a model that also includes both a model of galaxy formation and spin temperature  fluctuations. We find that the assumption of Gaussianity for the 21-cm field results in underestimating the cosmic variance of the 21-cm PS by a factor of $\gtrsim2$ for the scales relevant for the SKA ($k\sim 0.1-0.5$ Mpc$^{-1}$).

\section*{Acknowledgements}
We thank the referee for their detailed comments which improved the quality of this manuscript. This research was supported by the Australian Research Council Centre of Excellence for All Sky Astrophysics in 3 Dimensions (ASTRO 3D), through project \#CE170100013. Part of this work was performed on the OzSTAR national facility at the Swinburne University of Technology. The OzSTAR program partially receives funding from the Astronomy National Collaborative Research Infrastructure Strategy (NCRIS) allocation provided by the Australian Government.
This research was also undertaken with the assistance of resources from the National Computational Infrastructure (NCI Australia), an NCRIS-enabled capability supported by the Australian Government. YQ acknowledges ASTAC Large Programs and the Research Computing Services NCI Access scheme at The University of Melbourne for some NCI computing resources.

Software citations:
\begin{itemize}
    \item \textsc{PYTHON}: \cite{PYTHON}
    \item \textsc{NUMPY}: \cite{NUMPY}
    \item \textsc{SCIPY}: \cite{SCIPY}
    \item \textsc{MATPLOTLIB}: \cite{MATPLOTLIB}
    \item \textsc{CYTHON}: \cite{CYTHON}
    \item \textsc{CORRFUNC}: \cite{CORRFUNC}
\end{itemize}

\section*{Data Availability}
The data underlying this article will be shared on reasonable request to the corresponding author.

\bibliographystyle{mnras}
\bibliography{references}

\bsp
\label{lastpage}
\end{document}

%% file: new_commands.tex
\newcommand\lsim{\mathrel{\rlap{\lower4pt\hbox{\hskip1pt$\sim$}}
        \raise1pt\hbox{$<$}}}
\newcommand\gsim{\mathrel{\rlap{\lower4pt\hbox{\hskip1pt$\sim$}}
        \raise1pt\hbox{$>$}}}

\newcommand{\lya}{\ifmmode\mathrm{Ly}\alpha\else{}Ly$\alpha$\fi}
\newcommand{\lyb}{\ifmmode\mathrm{Ly}\beta\else{}Ly$\beta$\fi}
\newcommand{\igm}{\ifmmode\mathrm{IGM}\else{}IGM\fi}
\newcommand{\lae}{\ifmmode\mathrm{LAE}\else{}LAE\fi}
\newcommand{\h}{\ifmmode\mathrm{H}\else{}H\fi}
\newcommand{\hi}{\ifmmode\mathrm{H\,{\scriptscriptstyle I}}\else{}H\,{\scriptsize I}\fi}
\newcommand{\hii}{\ifmmode\mathrm{H\,{\scriptscriptstyle II}}\else{}H\,{\scriptsize II}\fi}
\newcommand{\cmb}{\ifmmode\mathrm{CMB}\else{}CMB\fi}
\newcommand{\eor}{\ifmmode\mathrm{EoR}\else{}EoR\fi}

\newcommand{\appropto}{\mathrel{\vcentre{
			\offinterlineskip\halign{\hfil$##$\cr
				\propto\cr\noalign{\kern2pt}\sim\cr\noalign{\kern-2pt}}}}}

\newcommand{\Rom}[1]{\uppercase\expandafter{\romannumeral #1}}
\newcommand{\rom}[1]{\lowercase\expandafter{\romannumeral #1}}

\newcommand{\hone}{\ifmmode\mathrm{\ion{H}{i} }\else{}\ion{H}{i} \fi}

\newcommand{\Msun}{\ifmmode\mathrm{M_\odot}\else{}$\rm M_\odot$\fi}

\newcommand{\oneh}{\text{ }h^{-1}}

\newcommand{\dforest}{\textsc{DarkForest} }

\newcommand{\genpsim}{\textsc{L210\_AUG} }
\newcommand{\mgenpsim}{\textsc{L210\_AUG\_lowX} }
\newcommand{\pgenpsim}{\textsc{L210\_AUG\_highX} }
\newcommand{\nrgenpsim}{\textsc{L210\_AUG\_nr} }
    
\newcommand{\gensim}{\textsc{L210} }
\newcommand{\nrgensim}{\textsc{L210\_nr} }

\newcommand{\xH}{\ifmmode\overline{x}_{\rm HI}\else{}$\overline{x}_{\rm HI}$\fi}

\newcommand{\cmfast}{\textsc{21cmFAST} }

\newcommand{\meraxes}{\ifmmode\mathrm{\textsc{Meraxes}}\else{}{\textsc{Meraxes} }\fi}